\documentclass[5p]{elsarticle}

\usepackage[UKenglish]{babel}
\usepackage[colorlinks]{hyperref}
\usepackage{subcaption}
\usepackage{amsmath,commath}
\DeclareMathOperator{\cov}{cov}
\DeclareMathOperator{\expectation}{E}

\usepackage{lineno}
\newcommand*\patchAmsMathEnvironmentForLineno[1]{
  \expandafter\let\csname old#1\expandafter\endcsname\csname #1\endcsname
  \expandafter\let\csname oldend#1\expandafter\endcsname\csname end#1\endcsname
  \renewenvironment{#1}%
     {\linenomath\csname old#1\endcsname}%
     {\csname oldend#1\endcsname\endlinenomath}}%
\newcommand*\patchBothAmsMathEnvironmentsForLineno[1]{%
  \patchAmsMathEnvironmentForLineno{#1}%
  \patchAmsMathEnvironmentForLineno{#1*}}%
\AtBeginDocument{%
\patchBothAmsMathEnvironmentsForLineno{equation}%
\patchBothAmsMathEnvironmentsForLineno{align}%
}

\usepackage{bm}
\renewcommand{\vec}[1]{\bm{#1}}

\journal{NIM A}

\begin{document}

    \begin{frontmatter}
        \title{Unfolding with Gaussian Processes}

        \author{Adam Bozson\corref{email}}
        \author{Glen Cowan\corref{none}}
        \author{Francesco Span\`o\corref{none}}
        \cortext[email]{Corresponding author: adam.bozson@cern.ch}
        \address{Department of Physics\\ Royal Holloway, University of London\\ Egham, Surrey, TW20~0EX, United Kingdom}      

        \begin{abstract}
            A method to perform unfolding with Gaussian processes (GPs) is presented.
            Using Bayesian regression, we define an estimator for the underlying truth distribution as the mode of the posterior.
            We show that in the case where the bin contents are distributed approximately according to a Gaussian, this estimator is equivalent to the mean function of a GP conditioned on the maximum likelihood estimator.
            Regularisation is introduced via the kernel function of the GP, which has a natural interpretation as the covariance of the underlying distribution.
            This novel approach allows for the regularisation to be informed by prior knowledge of the underlying distribution, and for it to be varied along the spectrum.
            In addition, the full statistical covariance matrix for the estimator is obtained as part of the result.
            The method is applied to two examples: a double-peaked bimodal distribution and a falling spectrum.
        \end{abstract}
    

    \end{frontmatter}


    \section{Introduction}
    Experimental measurements are distorted and biased by detector effects, due to limitations of the measuring instrument and procedures.
    The need to infer the underlying distribution using the measured data is shared by variety of fields, from astronomy \cite{Foreman-Mackey2017} and medical applications \cite{Andersen2002} to the investigation of the parameters that describe oil well properties \cite{Christen2016}.

    In most of these fields, these techniques are called \emph{deconvolution} or \emph{restoration} \cite{Hunt1977}.
    They are used to solve what is defined as the {\em inverse problem}: to infer an unknown function $f(\vec{x})$ from the measured data, using knowledge and assumptions of the distortions.

    In particle physics such techniques are known as \emph{unfolding} and a variety of methods have been developed for this purpose (for some reviews see Refs.~\cite{CowanUnfolding, Blobel,Spano:2013nca}).

    In this paper, a novel Bayesian method to perform unfolding in particle physics is proposed.
    We use an approach that ``gives prior probability to every possible function'' via \emph{Gaussian process regression} \cite{Rasmussen2006}, where higher probabilities are assigned to functions that are considered to agree with the observations.
    This approach allows greater flexibility than unfolding schemes based on a set of parametrised functions belonging to a specific class.
    In addition, it is shown to have a locally tunable regularisation scheme in terms of the variable to be unfolded.

    In Sec.~\ref{sec:unfolding}, we define the unfolding problem and the notation for approximately Gaussian-distributed datasets. 
    Sec.~\ref{sec:ML} discusses the solution to the unfolding problem based on the maximum likelihood (ML) method, and the need for regularisation.
    In a Bayesian setting, the likelihood is enhanced by prior information so that the ML solution is replaced by the mode of the posterior distribution.
    Sec.~\ref{sec:GP} connects the maximum \emph{a posteriori} (MAP) estimator to the solution of a regression problem which conditions prior knowledge encoded in a Gaussian process on the ML solution extracted from data.
    Example applications are provided in Sec.~\ref{sec:examples}.
    Finally, we report the conclusions and outlook for future exploration of this method in Sec.~\ref{sec:conclusionOut}.

    \section{Definitions and notation}\label{sec:unfolding}
    In particle physics, measured distributions are often reported as populations of bins rather than continuous functions.
    Therefore the first step we will take is to represent the underlying distributions with discretised bin populations.
    We note that this process biases the estimated histogram away from the true distribution.

    The truth distribution is referred to as $f(\vec{x}) $ and represented by a histogram $\vec{\mu} = (\mu_1, \ldots, \mu_M)$ with contents $\mu_j\propto\int_{\text{bin\,}j}f(\vec{x})\dif\vec{x}$, $j=1,\ldots,M$.
    Observed data are contained in a histogram $\vec{n} = (n_1, \ldots, n_N)$ with $N$ bins.
    The expectation value of $\vec{n}$ is the histogram $\vec{\nu}$.

    The truth and observed distributions are related through the effects of detector response, acceptance, and background contributions.
    For simplicity, we take the background to be zero (the relaxation of this assumption is discussed in Sec.~\ref{sec:conclusionOut}).
    The contents of $\vec{\mu}$ and $\vec{\nu}$ are linearly related by
    \begin{equation}
        \vec{\nu} = R \vec{\mu}, \label{eq:fold}
    \end{equation}
    where $R$ is the $N \times M $ \emph{response matrix} with elements $R_{ij}$ giving the conditional probability for an event to lie in bin $i$ of the observed histogram, given its true value is in bin $j$ of the truth histogram.
    
    The goal of unfolding is to construct estimators $\hat{\vec{\mu}}$ for the truth histogram $\vec{\mu}$, along with their covariance matrix $U_{ij} = \cov[\hat{\mu}_i,\hat{\mu}_j]$.
    In an experiment, the bin counts of the observed histogram $n_i$ fluctuate according to the Poisson distribution with expectation values $\nu_i$ and covariance matrix $V_{ij} = \nu_i \, \delta_{ij}$.
    The findings in this paper apply when bin counts are approximately Gaussian, i.e., for large $\nu_i$.

    \section{ML solution and regularisation}\label{sec:ML}
    Since the data $\vec{n}$ are approximately Gaussian-distributed around $\vec{\nu}$, the likelihood is given by
    \begin{equation}
        P(\vec{n}|\vec{\nu}) = \left[ (2\pi)^N \det(V) \right]^{-\frac{1}{2}} \exp\left[ -\frac{1}{2}(\vec{n} - \vec{\nu})^\mathsf{T} V^{-1} (\vec{n} - \vec{\nu}) \right]
    \end{equation}
    and hence the log-likelihood may be written
    \begin{equation}
        \log P(\vec{n}|\vec{\mu}) = -\frac{1}{2} \left( \vec{n} - R\vec{\mu} \right)^\mathsf{T} V^{-1} \left( \vec{n} - R\vec{\mu} \right) + \ldots, \label{eq:likelihood}
    \end{equation}
    where we have substituted $\vec{\nu} = R\vec{\mu} $ and not written terms that do not depend on $ \vec\mu $.
    It can be shown that the maximum likelihood (ML) solution $\hat{\vec{\mu}}_\text{ML}$ satisfies $\vec{n} = R\hat{\vec{\mu}}_\text{ML}$.
    We write the ML solution as $\hat{\vec{\mu}}_\text{ML} = R^{-1} \vec{n}$, which may be obtained by explicit matrix inversion for invertible $R$ when $N=M$ or by alternative methods, such as numerically maximising Eq.~\eqref{eq:likelihood} or singular value decomposition.
    The ML covariance matrix is given by $U_\text{ML} = R^{-1} V \left( R^{-1} \right)^\textsf{T}$ \cite{Cowan1998}.

    The detector response acts to smear out fine structure in the truth distribution, so statistical fluctuations in the data can lead to a large amount of fine structure in the unfolded result.
    This effect yields large local fluctuations in the ML unfolded solution when the typical bin width is not much larger than the detector resolution.
    In addition, the estimators for neighbouring bin counts can often have strong negative correlations.

    These unwanted false features are typically reduced by a technique known as \emph{regularisation}.
    Regularisation may be introduced by minimising a \emph{cost functional} $\Phi(\vec{\mu}) = -\alpha \log P(\vec{n}|\vec{\mu}) + S(\vec{\mu}) $, where $S(\vec{\mu})$ penalises high-variance distributions, effectively constricting the space of possible unfolded solutions.
    Multiple measures of smoothness may be used, such as those based on derivatives \cite{Tikhonov1952,Phillips1962} or entropy \cite{Schmelling1994}.
    The ML solution has the minimum variance for an unbiased estimator, so any reduction in variance must be balanced by introducing some bias.
    The regularisation parameter $\alpha$ controls this bias--variance trade-off.

    An unfolded distribution may alternatively be obtained by iterative techniques \cite{Zech2013,DAgostini1995}, which converge on the ML solution.
    Stopping after a fixed number of iterations can yield a solution with the desired properties, although the fact that the bias--variance trade-off is controlled by a discrete parameter, rather than a continuous one, is seen as a disadvantage because it limits the possibility to tune the parameter values.
    \emph{Fully Bayesian unfolding} \cite{Choudalakis2012} addresses regularisation through a non-constant prior distribution, and performs the unfolding by sampling from the posterior distribution.

    \section{Gaussian process method}\label{sec:GP}
    The method presented in this paper builds on the ML solution from Sec.~\ref{sec:ML}.
    Starting from the Gaussian likelihood given by Eq.~\eqref{eq:likelihood} and a \emph{Gaussian process} (GP) prior, Bayes' theorem is applied to obtain a posterior distribution.
    We define the estimator representing the unfolded distribution as a summary statistic of the posterior, namely the mode.
    We remark that while the ML solution is a frequentist estimator, the method presented here incorporates elements of Bayesian statistics.
    However the final estimator for the unfolded distribution is a valid frequentist estimator, so this method may be used in either fully Bayesian or hybrid analyses.

    A \emph{random process} extends the notion of a random variable to the space of functions of a set of indices $X$.
    A GP is therefore a set of indexed random variables, any finite subset of which are distributed according to a joint Gaussian distribution \cite{Rasmussen2006}.
    Since the Gaussian distribution is entirely defined by its mean and covariance, a complete description of a GP requires just a mean function $m(\vec{x}) = \expectation[f(\vec{x})]$ and a \emph{kernel function} $k(\vec{x}, \vec{x}^\prime) = \cov[f(\vec{x}),f(\vec{x}^\prime)]$.
    A GP can be thought of as a probability distribution for the \emph{latent function} $f(\vec{x})$.

    GPs may be used for regression, where one wishes to estimate the function $f$ of the (generally multidimensional) variable $\vec{x}$, given some observations $\vec{y}$ taken at $X = (\vec{x}_1, \allowbreak\vec{x}_2, \ldots)$.
    This is done by updating a GP prior using Bayes' theorem to obtain a posterior GP for $f$.
    The mean (or equivalently the mode) of the posterior, evaluated at $X_* = (\vec{x}_{*1}, \vec{x}_{*2}, \ldots)$, is denoted $\bar{\vec{f}}_*$ and used as the estimator for $f$.
    A rich treatment of using GPs for regression may be found in Ch.~2 of Ref.~\cite{Rasmussen2006}, whose notation we follow in this paper.
    Here we state the posterior mean and covariance of a GP with prior mean function $m$ and kernel function $k$ for a vector of observations $\vec{y}$ with data covariance matrix $V$:
    \begin{align}
        \bar{\vec{f}}_* &= K_*^\mathsf{T} \left[ K + V \right]^{-1} \vec{y} + \vec{m}_* ,  \label{eq:GP-mean} \\
        \cov(\vec{f}_*) &= K_{**} - K_*^\mathsf{T} \left[ K + V \right]^{-1} K_* , \label{eq:GP-cov}
    \end{align}
    with the matrices $K_{ij}=k(\vec{x}_i,\vec{x}_j)$, $[K_*]_{ij} = k(\vec{x}_i, \vec{x}_{*j})$, $[K_{**}]_{ij} = k(\vec{x}_{*i}, \vec{x}_{*j})$.
    Here $\vec{m}_* = m(X_*)$.

    This standard result from GP regression is used in the following section to link the estimator for an unfolded distribution to a GP.

    \subsection{MAP estimator}\label{sec:MAP-est}
    We consider again the model from Sec.~\ref{sec:ML} with the likelihood $P(\vec{n}|\vec{\mu})$ given by Eq.~\eqref{eq:likelihood}.
    From Bayes' theorem, the log-posterior probability is given by
    \begin{equation}
        \log P(\vec{\mu}|\vec{n}) = \log P(\vec{n} | \vec{\mu}) + \log P(\vec{\mu}) - \log P(\vec{n}), \label{eq:posterior}
    \end{equation}
    where $P(\vec{\mu})$ is the prior probability, and the last term $P(\vec{n})$ may be ignored since it does not depend on $\vec{\mu}$.

    We take the prior probability to be given by a GP with mean vector $\vec{m}$ (the \emph{reference histogram}) and covariance matrix $K_{ij} = k(\vec{x}_i, \vec{x}_j)$.
    The log-prior probability is then given by
    \begin{equation}
        \log P(\vec{\mu}) = -\frac{1}{2}\left( \vec{\mu} - \vec{m} \right)^\mathsf{T} K^{-1} \left( \vec{\mu} - \vec{m} \right) + \ldots. \label{eq:prior}
    \end{equation}
    Substituting the likelihood Eq.~\eqref{eq:likelihood} and prior Eq.~\eqref{eq:prior} into the expression for the posterior Eq.~\eqref{eq:posterior}, we obtain
    \begin{align}
        \log P(\vec{\mu}|\vec{n}) =& -\frac{1}{2} \left( \vec{n} - R\vec{\mu} \right)^\mathsf{T} V^{-1} \left( \vec{n} - R\vec{\mu} \right) \nonumber \\ 
        & - \frac{1}{2}\left( \vec{\mu} - \vec{m} \right)^\mathsf{T} K^{-1} \left( \vec{\mu} - \vec{m} \right) + \ldots, \label{eq:MAPposterior}
    \end{align}
    dropping terms which do not contain $\vec{\mu}$.
    The maximum \textit{a posteriori} (MAP) estimator $\hat{\vec{\mu}}$ is the mode of this posterior probability, and maximises $\log P(\vec{\mu} | \vec{n})$.
    This summary statistic is found to be given by
    \begin{equation}
        \hat{\vec{\mu}} = K \left[ K + R^{-1}V(R^{-1})^\mathsf{T} \right]^{-1} \left( R^{-1}\vec{n} - \vec{m} \right) + \vec{m} \label{eq:MAP}.
    \end{equation}
    A derivation of this is given in \ref{sec:MAP}.

    By comparing the MAP estimator from Eq.~\eqref{eq:MAP} to that obtained from GP regression in Eq.~\eqref{eq:GP-mean}, we find the important result that $\hat{\vec{\mu}}$ is the posterior mean of a GP regression whose observations are the ML solution, which is given by $\hat{\vec{\mu}}_\text{ML} = R^{-1}\vec{n} $ with covariance matrix $U_\text{ML} = R^{-1}V(R^{-1})^\mathsf{T}$.
    Since the posterior distribution is a product of Gaussians, it is also Gaussian and therefore the mode is identical to the mean.
    This connection allows us to write that the covariance of the MAP estimator may be given by
    \begin{equation}
        U = K - K\left[K + R^{-1}V(R^{-1})^\mathsf{T}\right]^{-1}K. \label{eq:MAP-var}
    \end{equation}

    Furthermore, if the observation (training) indices $X = (\vec{x}_1, \vec{x}_2, \ldots)$ are different from the prediction (testing) indices $X_* = (\vec{x}_{*1}, \vec{x}_{*2}, \ldots)$, and the reference histogram can be obtained for bins defined by $X_*$, then we may use the standard results from GP regression to generalise the MAP solution to
    \begin{align}
        \hat{\vec{\mu}} &= K_*^\mathsf{T}\left[ K + U_\text{ML} \right]^{-1}\left( \hat{\vec{\mu}}_\text{ML} - \vec{m} \right) + \vec{m}_*, \label{eq:predictive-mean}\\
        U &= K_{**} - K_*^\mathsf{T}\left[ K + U_\text{ML} \right]^{-1}K_*, \label{eq:predictive-covariance}
    \end{align}
    where $[K_{*}]_{ij} = k(\vec{x}_i, \vec{x}_{*j})$, $[K_{**}]_{ij} = k(\vec{x}_{*i}, \vec{x}_{*j})$, and $\vec{m}_*$ is the mean histogram at $X_*$.

    The generalised results in Eq.~\eqref{eq:predictive-mean} and Eq.~\eqref{eq:predictive-covariance} are simple algebraic expressions once the ML solution is known.
    Therefore the unfolded estimator and covariance are efficient to compute.
    This is an advantage over other, more CPU-intensive unfolding schemes.
    In addition, these results are linear in $\vec{n}$ so error propagation is simple.

    \subsection{Kernel choice and optimisation}\label{sec:opt}
    In the proposed GP unfolding method, the regularisation is introduced via the kernel function $k(\vec{x},\vec{x}^\prime)$ which constricts the space of possible solutions to those with a particular covariance.
    A common choice for the kernel function is the squared-exponential:
    \begin{equation}
        k(\vec{x},\vec{x}^\prime) = A \, \exp\left(\frac{-\norm{\vec{x}-\vec{x}^\prime}^2}{2l^2}\right). \label{eq:sq-exp}
    \end{equation}
    This kernel function is \emph{stationary} in the sense that it is a function of only the distance between the inputs, $\norm{\vec{x}-\vec{x}^\prime}$.
    It is parameterised by the amplitude $A$ and length scale $l$, referred to as the set of \emph{hyperparameters}, $\vec{\theta} = \{A, l\}$.
    Various methods exist to choose their values.
    
    One method for this is by simulation, as is often done in particle physics analyses.
    In this approach, a simulation program produces values for $\vec{\mu}$, $\vec{\nu}$, and statistically independent $\vec{n}$.
    Then the pseudo-data $\vec{n}$ are unfolded with varying hyperparameters to obtain $\hat{\vec{\mu}}_{\vec{\theta}}$, and the agreement with $\vec{\mu}$ checked for \emph{closure}.
    The acceptable degree of closure, and its measure, are often chosen by eye, although a more specific goodness-of-fit statistic may be used.

    Another approach, taken from GP methods \cite{Rasmussen2006}, is to maximise the marginal likelihood,
    \begin{align}
        \log P(\vec{n};\vec{\theta}) =& \log \left( \int P(\vec{n}|\vec{\mu}) \, P(\vec{\mu};\vec{\theta}) \, \dif{\vec{\mu}} \right) \nonumber \\
        =& -\frac{1}{2}\left( \hat{\vec{\mu}}_\text{ML} - \vec{m} \right)^\mathsf{T} \left[ K_{\vec{\theta}} + U_\text{ML} \right]^{-1} \left( \hat{\vec{\mu}}_\text{ML} - \vec{m} \right) \nonumber \\
        &-\frac{1}{2}\log|{K_{\vec{\theta}} + U_\text{ML}}| - \frac{N}{2}\log{2\pi}, \label{eq:marginal-lh}
    \end{align}
    where $K_{\vec{\theta}}\ = k_{\vec{\theta}}(X,X)$ is the kernel function evaluated at $X$ with the hyperparameters set to $\vec{\theta}$.
    This is a Bayesian approach, marginalising over the latent distribution $\vec{\mu}$.
    The maximum of this marginal likelihood defines a model with a trade-off between the fit to the data (the first term) versus model complexity (the second term).
    For example, a GP using a squared-exponential kernel with very small length scale $l$ will tend to fit to $\hat{\vec{\mu}}_\text{ML}$, but will be overly complex (under-regularisation).
    In contrast, a large $l$ describes a simpler model, but will fail to fit to the $\hat{\vec{\mu}}_\text{ML}$ (over-regularisation).
    These extreme situations are penalised by the marginal likelihood, whose maximum point may be used to choose values for $\vec{\theta}$.

    Finally, we mention the method of \emph{cross validation}, often employed for hyperparameter optimisation in machine learning \cite{scikit-learn}.
    Various approaches for cross validation exist, but given the relatively small number of bins $M$ in a typical unfolding scenario, and the fast computation of the GP unfolded result, we recommend the \emph{leave-one-out} variant.
    Here, $M$ sets of $M-1$ bins, $X_*^{(1)}, X_*^{(2)}, \ldots, X_*^{(M)}$, are produced with each $X_*^{(i)}$ missing the $i$th bin.
    Then the prediction for $\mu_i$ is compared to its true value via a \emph{loss function}, most often the squared error.
    The set of $M$ losses for some hyperparameters $\vec{\theta}$ can then be used to choose their values.

    Other kernel functions may be more suitable for describing the truth distribution.
    An attractive feature of the approach presented here is that one may encode knowledge of the underlying physical process to derive a physically-motivated kernel \cite{Frate2017} which may better describe the truth distribution.
    
    The mathematics of \emph{reproducing kernel Hilbert spaces} formalises the link between the kernel and the traditional regularisation approach used in some particle physics results.
    For example, a \emph{thin plate covariance} \cite{Williams2007} leads to a solution equivalent to that of spline regularisation, known as Tikhonov regularisation in particle physics \cite{Cowan1998,Tikhonov1952,Phillips1962,Wahba1990}.
    In one dimension, this stationary kernel may be written $k(r) = A (2r^3 - 3Rr^2 + R^3)$, where $r=|x-x^\prime|$ and $R$ is determined by boundary conditions.
    This kernel contains a single parameter \(A\), which controls the global strength of the regularisation, as is the case with Tikhonov regularisation in its usual implementation.
    In contrast, an advantage of the GP approach presented in this paper is that the regularisation may be varied locally along the spectrum by using a non-stationary kernel function.
    We provide an example of this in Sec.~\ref{sec:falling-spectrum}.

    \section{Example applications}\label{sec:examples}
    Python code for the following examples may be found in Ref.~\cite{github}.
    We consider the case of a bimodal distribution in Sec.~\ref{sec:bimodal}, and a falling spectrum in Sec.~\ref{sec:falling-spectrum}.

    \subsection{Bimodal distribution}\label{sec:bimodal}
\begin{figure*}
    \begin{minipage}{.48\textwidth}
        \centering
        \includegraphics[width=\textwidth]{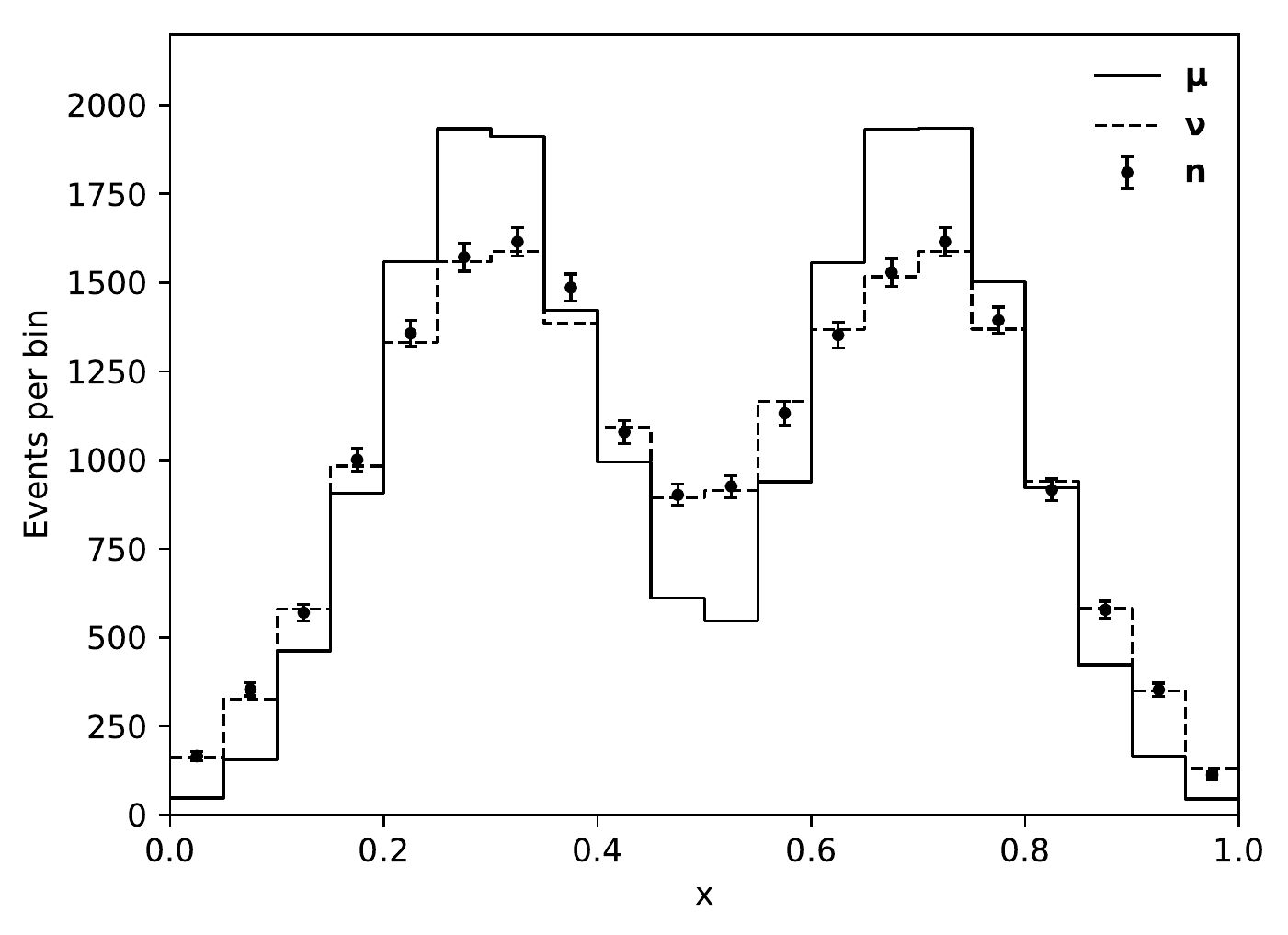}
        \caption{Truth ($\vec{\mu}$), expected ($\vec{\nu}$), and observed ($\vec{n}$) histograms for the two-peak unfolding example.
            The histogram definitions are reported in the text.
            The error bars on $\vec{n}$ represent their Poisson uncertainties.\label{fig:ingredients}}
    \end{minipage}\hfill
    \begin{minipage}{.48\textwidth}
        \centering
        \includegraphics[width=\textwidth]{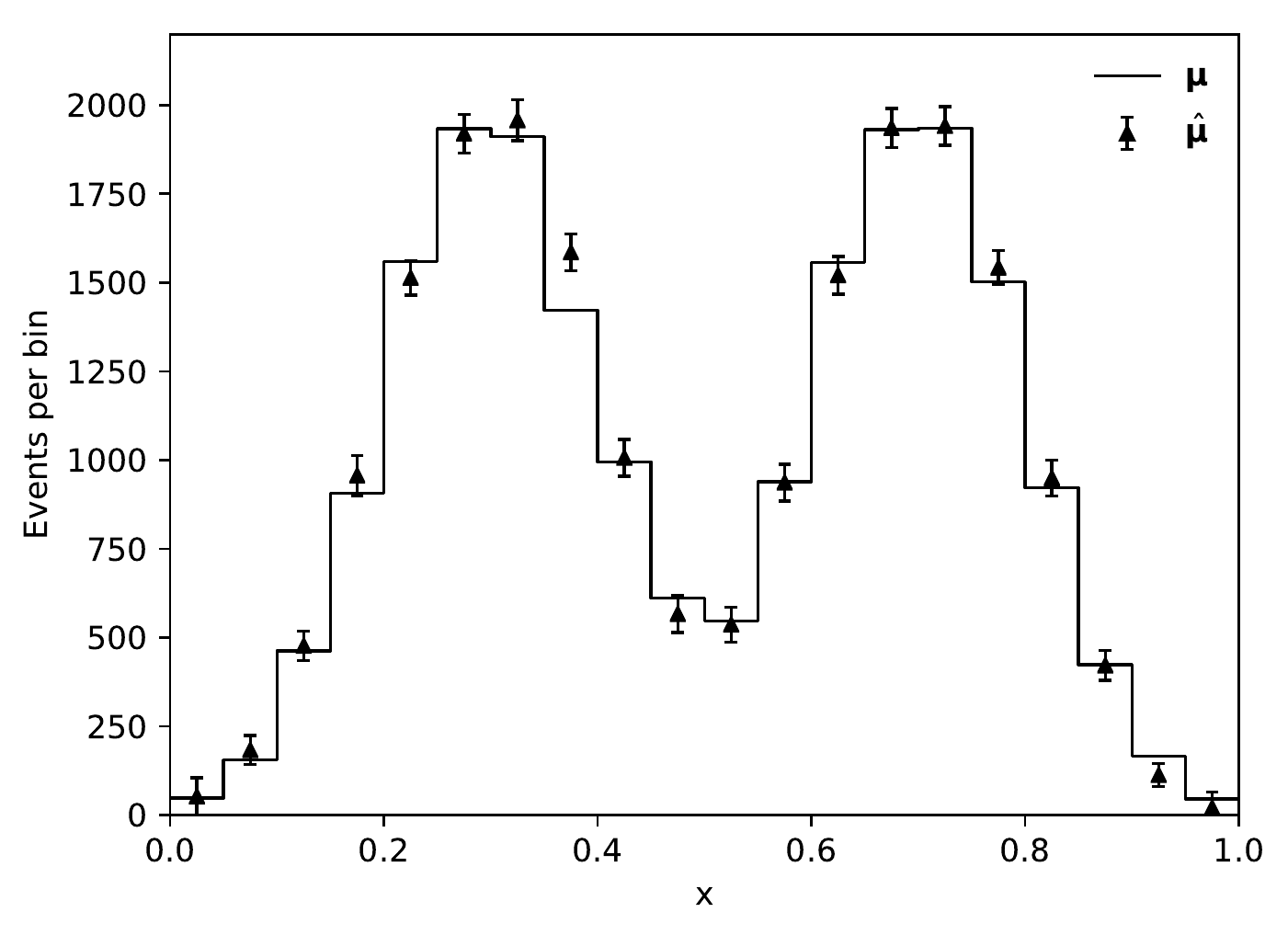}
        \caption{Truth ($\vec{\mu}$) and unfolded truth estimators ($\hat{\vec{\mu}}$) for the two-peak example.
        The error bars on $\hat{\vec{\mu}}$ represent the standard deviations obtained from the covariance matrix as defined by Eq.~\eqref{eq:MAP-var}.\label{fig:result}\vspace{1em}}
    \end{minipage}
    \end{figure*}

    A set of 20\,000 toy truth events is obtained by sampling from two Gaussian distributions for $x$ with mean values 0.3 and 0.7, both with standard deviation 0.1.
    These truth events are histogrammed in $\vec{\mu}$.
    The truth events are then smeared with a Gaussian resolution of $\sigma=0.075$ to generate the histogram $\vec{\nu}$.
    Events are accepted in the region $0 < x < 1$ and both the $\vec{\mu}$ and $\vec{\nu}$ histograms use 20 constant-width bins.
    The truth and smeared events are used to determine the response matrix $R$ from a normalised 2D histogram.
    Finally, the observed histogram $\vec{n}$ is generated by Poisson fluctuations around $\vec{\nu}$.
    The three histograms are shown in Fig.~\ref{fig:ingredients}.

\begin{figure}
        \centering
        \includegraphics[width=0.5\textwidth]{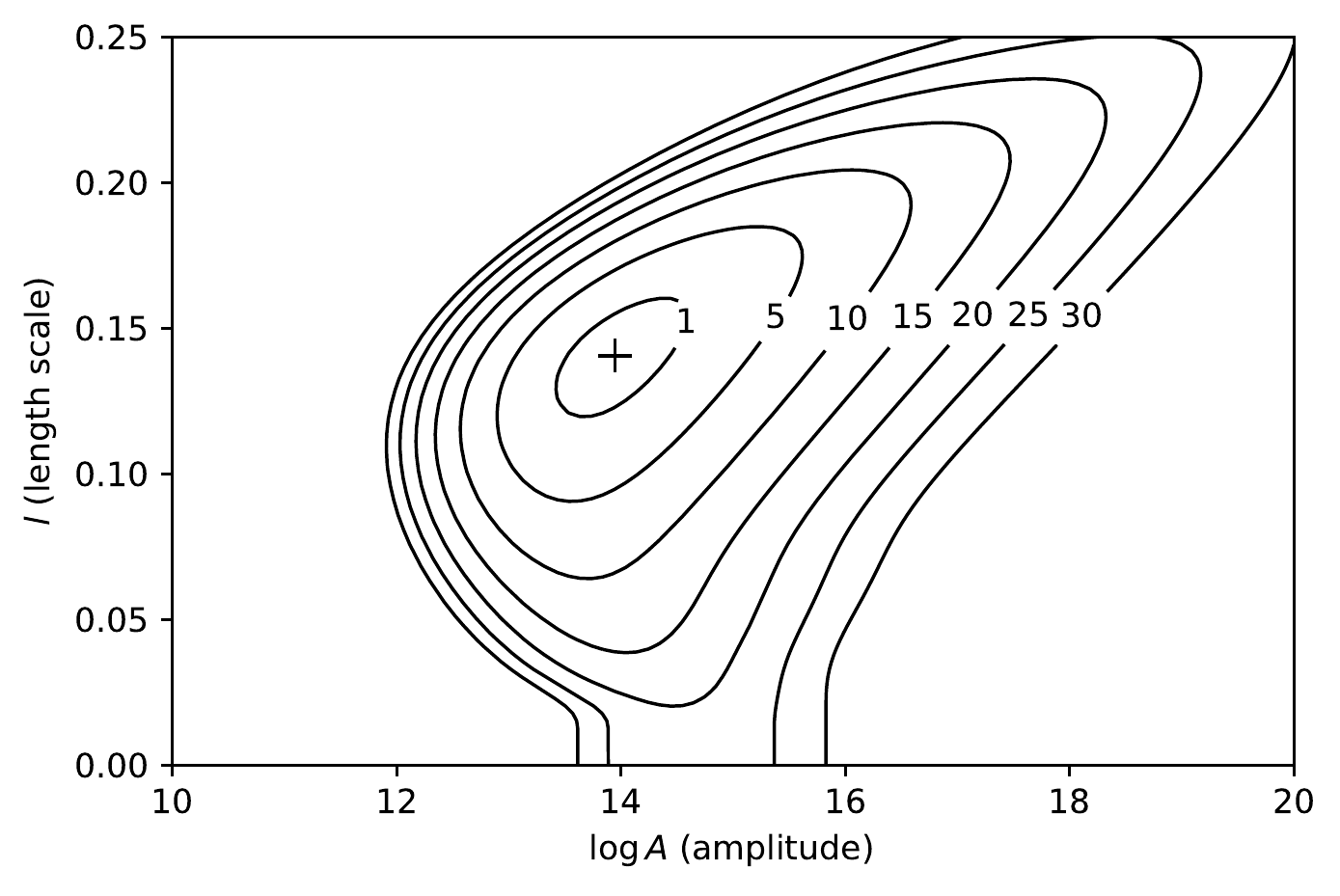}
        \caption{Contours of the log-marginal likelihood Eq.~\eqref{eq:marginal-lh} for the two-peak example as a function of the parameters for the squared-exponential kernel, $A$ and $l$.
            The cross indicates the point of maximum marginal likelihood.
            The contour labels are the depth of the contour below the maximum.\label{fig:marginal-lh}}
    \end{figure}

\begin{figure}
        \centering
        \includegraphics[width=0.5\textwidth]{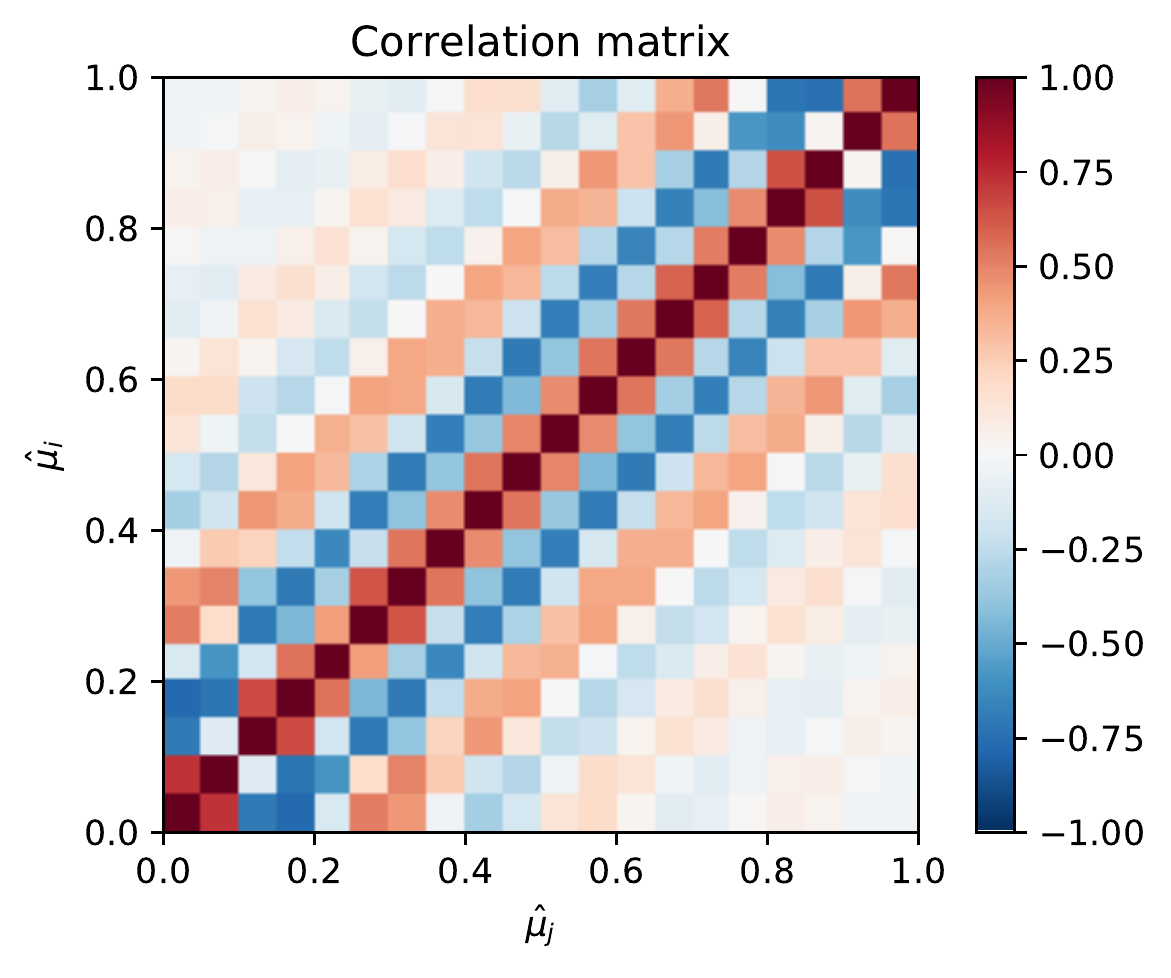}
        \caption{Correlation matrix for the unfolded truth estimators $\hat{\vec{\mu}}$ for the two-peak example.\label{fig:result-cov}}
    \end{figure}

    We use a GP with the squared-exponential kernel function given by Eq.~\eqref{eq:sq-exp}.
    The values for the two hyperparameters $A$ and $l$ are chosen to be those that maximise the marginal likelihood given by Eq.~\eqref{eq:marginal-lh}.
    The maximum and contours of the marginal likelihood are shown in Fig.~\ref{fig:marginal-lh}.
    The mean histogram for the GP, $\vec{m}$, is taken to be 0 for all bins since it is found to have little impact on the final result.
    The unfolded estimator for the truth, $\hat{\vec{\mu}}$ given by Eq.~\eqref{eq:MAP}, is shown in Fig.~\ref{fig:result}.
    The covariance matrix $U$ is defined by Eq.~\eqref{eq:MAP-var}, and the correlation matrix $\rho_{ij} = U_{ij}/\sqrt{U_{ii}U_{jj}}$ is shown in Fig.~\ref{fig:result-cov}.

    \subsection{Falling spectrum}\label{sec:falling-spectrum}
\begin{figure*}
    \begin{minipage}{.48\textwidth}
        \centering
        \includegraphics[width=\textwidth]{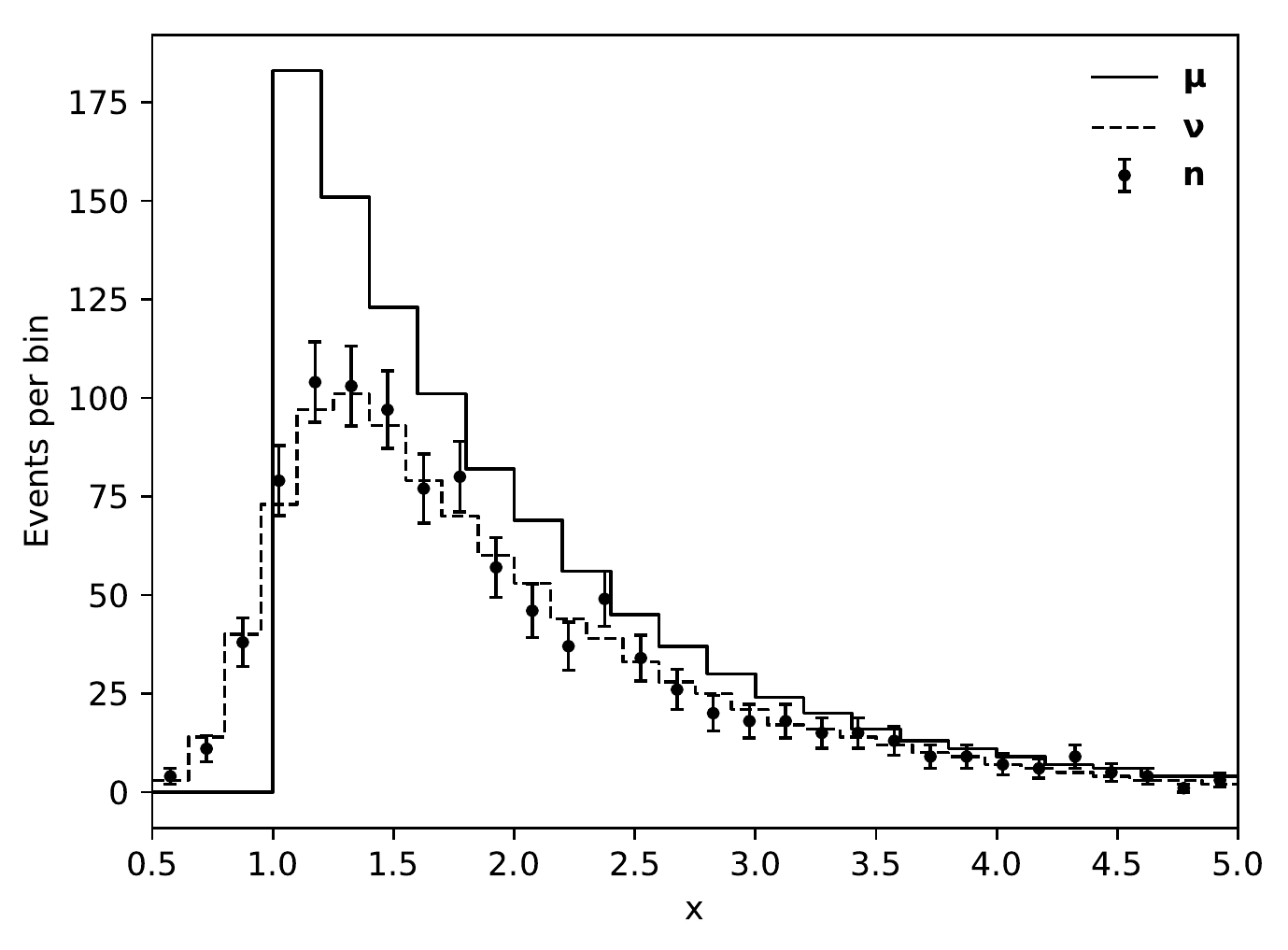}
        \caption{Truth ($\vec{\mu}$), expected ($\vec{\nu}$), and observed ($\vec{n}$) histograms for the falling spectrum example, as defined in the text.
            The error bars on $\vec{n}$ represent their Poisson uncertainties.\label{fig:ingredients-exp}}
    \end{minipage}\hfill
    \begin{minipage}{.48\textwidth}
        \centering
        \includegraphics[width=\textwidth]{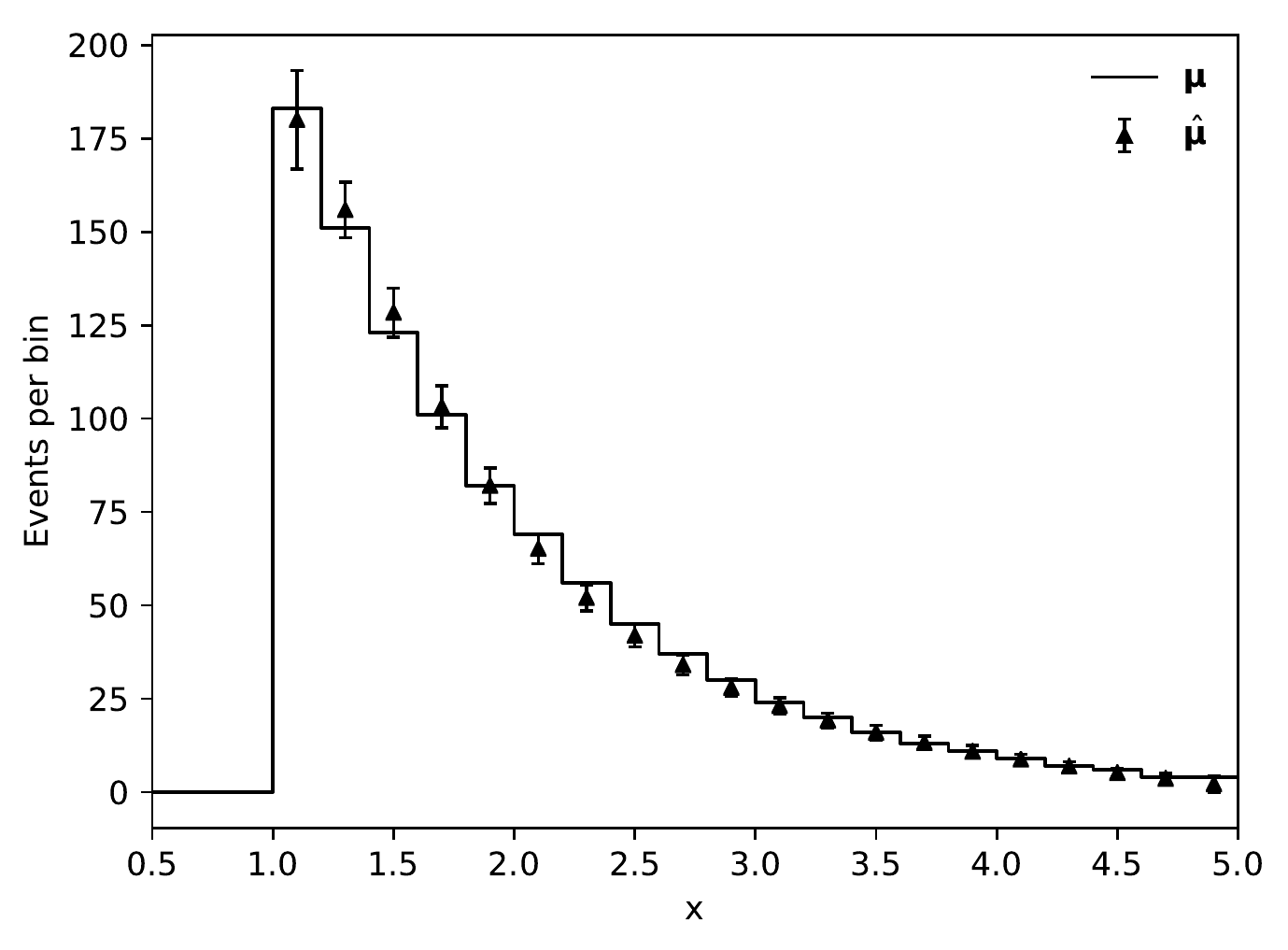}
        \caption{Truth ($\vec{\mu}$) and unfolded truth estimators ($\hat{\vec{\mu}}$) for the falling spectrum example.\vspace{1em}\label{fig:result-exp}}  
    \end{minipage}
    \end{figure*}

    1\,000 truth events are sampled from an exponential distribution $f(x) = e^{-x}$ in the region $1 < x < 5$ and accumulated in 20 bins of equal width ($\vec{\mu}$).
    These events are smeared according to a Gaussian with resolution $0.2\sqrt{x}$.
    The smeared events are placed in a histogram  $\vec{\nu}$ with 30 bins of equal width in the region $0.5 < x < 5$, and the observed histogram $\vec{n}$ is generated from a Poisson distribution around $\vec{\nu}$.
    These three histograms are shown in Fig.~\ref{fig:ingredients-exp}.
    
    In this example, $N > M$ so while the problem is well-constrained, the $N \times M$ response matrix $R$ is not directly invertible.
    To mitigate this, we use the ML estimator as the starting point and then regularise with the MAP prescription detailed in Sec.~\ref{sec:GP}.
    Specifically, we numerically maximise the Gaussian likelihood in Eq.~\eqref{eq:likelihood} using MINUIT \cite{1975CoPhC..10..343J} via the iminuit \cite{iminuit} Python interface to obtain the ML estimator $\hat{\vec{\mu}}_\text{ML}$.
    The covariance matrix for the ML estimator, $U_\text{ML}$ is obtained by inverting the Hessian matrix with the HESSE subroutine.
    These results are then substituted in Eq.~\eqref{eq:MAP} with $R^{-1}\vec{n} \rightarrow \hat{\vec{\mu}}_\text{ML}$ and $R^{-1}V(R^{-1})^\mathsf{T} \rightarrow U_\text{ML}$.

    Since the (\emph{a priori} known) detector resolution increases proportionally with $\sqrt{x}$, a kernel with constant length scale, such as the squared-exponential Eq.~\eqref{eq:sq-exp} is unsuitable in this case.
    Therefore we choose a kernel function with variable length scale, the Gibbs kernel \cite{Rasmussen2006, Gibbs1997} in 1D,
    \begin{equation}
        k(x, x^\prime) = A \, \sqrt{\frac{2 l(x) l(x^\prime)}{l^2(x) + l^2(x^\prime)}} \, \exp\left( -\frac{(x - x^\prime)^2}{l^2(x) + l^2(x^\prime)} \right),
    \end{equation}
    where $l(x)$ is an arbitrary positive function of $x$, here chosen to be $l(x) = bx + c$.
    This allows for a linearly-changing length scale.
    The increased flexibility afforded by this kernel function is realised by introducing more regularisation parameters, $\vec{\theta} = \{ A, b, c \}$.
    We remark that for a large number of parameters, it becomes increasingly difficult to choose the optimal point.

    The unfolded estimators are shown against the truth histogram in Fig.~\ref{fig:result-exp}.
    Here the parameters $\vec{\theta}$ are chosen with the maximum marginal likelihood prescription given in Sec.~\ref{sec:opt}.
    As expected, $b > 0$ so the length scale increases with $x$.

    \section{Conclusions and Outlook}\label{sec:conclusionOut}
    In this paper we have presented how GPs may be applied to the unfolding problem.
    It is shown that conditioning a GP prior on the ML solution is equivalent to constructing the MAP estimator.
    In this application, the use of a GP regressor may be thought of as a method of regularising the ML solution.

    The GP is entirely described by mean and kernel functions.
    While the mean function is found to have little impact on the result, the kernel function prescribes the covariance of unfolded estimation for the truth distribution.
    By choosing an appropriate kernel function, the smoothness in the unfolded estimator can be controlled.
    Furthermore, the kernel function has a direct interpretation and may be motivated by knowledge of the underlying physics.
    This means that, in contrast to other unfolding schemes, the regularisation is a natural product of this approach.

    For $N=M$, where the bins for the truth and observed histograms are equal, the ML solution is simply given by inverting the response matrix, $\hat{\vec{\mu}}_\text{ML} = R^{-1}\vec{n}$.
    However, generally $N \neq M$ and this method may not be used.
    For the general case we envisage two possibilities.
    First, the $N \times M$ response matrix $R$ is constructed from Monte Carlo simulation, and the ML solution is found numerically by maximising the likelihood given by Eq.~\eqref{eq:likelihood}.
    This is the approach taken in Sec.~\ref{sec:falling-spectrum} for the falling spectrum example.
    Alternatively, the square $N \times N$ matrix $R^\prime$ could be constructed using the same binning for the truth and observed histograms.
    Then the predictive GP mean function Eq.~\eqref{eq:predictive-mean} is evaluated at $X_*$, the $M$ centers of the bins for the desired truth histograms.

    In Sec.~\ref{sec:unfolding}, we take the background contribution to be equal to zero for simplicity.
    Background contributions, in the form of the $N$-dimensional vector $\vec{\beta}$, are simple to include by modifying the folding equation Eq.~\eqref{eq:fold} to $\vec{\nu} = R\vec{\mu} + \vec{\beta}$.
    Then for the estimators used in the method presented in this paper, one simply substitutes the data for the background-subtracted data, $\vec{n} \rightarrow \vec{n} - \vec{\beta}$.

    This paper assumes throughout that the data may be approximated as distributed according to a Gaussian, and we note that this is not universally the case in particle physics.
    However, the choice of unfolding method depends on the analysis being done and should be tested against simulation in any case.
    Therefore we recommend that for histograms with small bin populations, the unfolding is tested to ensure it acceptably meets the requirements of the analysis under consideration.

    The treatment of systematic uncertainties is postponed to future work in this area. 
    We remark that that approximate variational approaches, as used in published particle physics analyses \cite{Aad2016,Aad2016-2}, may still be employed in this case.
    We envisage further research into the applications of Student-$t$ processes \cite{shah2014student} in unfolding in particle physics, as an extension to this work.

    GPs have been introduced to a number of scientific fields to improve their statistical procedures \cite{Christen2016, Hunt1977}.
    They have not, however, traditionally been used in particle physics, although recent developments in this area have shown promise \cite{Frate2017}.
    In this paper, we have introduced GPs into the important problem of unfolding.
    We show that the method is generally applicable to problems of different shapes and sizes, that the regularisation can be controlled naturally, and that the result -- including the unfolded covariance matrix -- can be obtained conveniently.

    \section*{Acknowledgements}
    We would like to extend our gratitude to our colleagues in ATLAS and RHUL for their support.
    In particular, we thank Pedro Teixeira-Dias, and Lewis Wilkins for their proofreading and helpful comments on this manuscript.
    We also thank Veronique Boisvert and Pim Verschuuren for insightful discussions.
    This work was supported by the UK Science and Technology Facilities Council.

    \appendix
    \def\appendixname{Appendix } 
    \section{Derivation of MAP estimator}\label{sec:MAP}
    With reference to Eq.~\eqref{eq:MAPposterior}, we wish to find the value $\hat{\vec{\mu}}$ that maximises the expression
    \begin{align}
       -\frac{1}{2} \left( \vec{n} - R\vec{\mu} \right)^\mathsf{T} V^{-1} \left( \vec{n} - R\vec{\mu} \right) - \frac{1}{2}\left( \vec{\mu} - \vec{m} \right)^\mathsf{T} K^{-1} \left( \vec{\mu} - \vec{m} \right).
    \end{align}
    The derivative for each term is given by
    \begin{align}
        \frac{\partial}{\partial\vec\mu}\left[ -\frac{1}{2} \left( \vec{n} - R\vec{\mu} \right)^\mathsf{T} V^{-1} \left( \vec{n} - R\vec{\mu} \right) \right] &= \left( \vec{n} - R\vec{\mu} \right)^\mathsf{T} V^{-1} R, \\
        \frac{\partial}{\partial\vec\mu}\left[ - \frac{1}{2}\left( \vec{\mu} - \vec{m} \right)^\mathsf{T} K^{-1} \left( \vec{\mu} - \vec{m} \right) \right] &= -\left( \vec{\mu} - \vec{m} \right)^\mathsf{T} K^{-1}.
    \end{align}
    Combining these and taking the transpose ($V^{-1}$ and $K^{-1}$ are symmetric), we therefore require that $\hat{\vec{\mu}}$ satisfies
    \begin{align}
        \vec{0} &= R^\mathsf{T} V^{-1} \left( \vec{n} - R\hat{\vec{\mu}} \right) - K^{-1} \left( \hat{\vec{\mu}} - \vec{m} \right) \\
        &= R^\mathsf{T} V^{-1} \vec{n} - \left[ R^\mathsf{T} V^{-1} R + K^{-1} \right] \hat{\vec{\mu}} + K^{-1} \vec{m}. \label{eq:A5}
    \end{align}
    Now we use that the covariance of the ML solution from Sec.\ \ref{sec:ML} is given by $U_\text{ML} = R^{-1} V (R^{-1})^\mathsf{T} $ and therefore that $R^\mathsf{T} V^{-1} R = U_\text{ML}^{-1}$.
    Substituting into Eq.~\eqref{eq:A5} and rearranging for $\hat{\vec{\mu}}$,
    \begin{align}
        \hat{\vec{\mu}} &= \left[ K^{-1} + U_\text{ML}^{-1} \right]^{-1} \left( U_\text{ML}^{-1} R^{-1} \vec{n} + K^{-1} \vec{m} \right) \label{eq:A6}\\
        &= K \left[ K + U_\text{ML} \right]^{-1} R^{-1} \vec{n} + U_\text{ML} \left[ K + U_\text{ML} \right]^{-1} \vec{m} \label{eq:A7} \\
        &= K \left[ K + U_\text{ML} \right]^{-1} \left( R^{-1}\vec{n} - \vec{m} \right) + \vec{m},
    \end{align}
    where from Eq.~\eqref{eq:A6} to Eq.~\eqref{eq:A7} we use that \\ 
    $\left[ A^{-1} + B^{-1} \right]^{-1} B^{-1} \equiv A\left[ A+B \right]^{-1} $ for invertible matrices $A$ and $B$.

    \bibliography{bib}

\begin{thebibliography}{10}
\expandafter\ifx\csname url\endcsname\relax
  \def\url#1{\texttt{#1}}\fi
\expandafter\ifx\csname urlprefix\endcsname\relax\def\urlprefix{URL }\fi
\expandafter\ifx\csname href\endcsname\relax
  \def\href#1#2{#2} \def\path#1{#1}\fi

\bibitem{Foreman-Mackey2017}
D.~Foreman-Mackey, E.~Agol, S.~Ambikasaran, R.~Angus, Fast and scalable
  {Gaussian} process modeling with applications to astronomical time series,
  Astron. J. 154~(6) (2017) 220.
\newblock \href {http://arxiv.org/abs/1703.09710} {\path{arXiv:1703.09710}},
  \href {http://dx.doi.org/10.3847/1538-3881/aa9332}
  {\path{doi:10.3847/1538-3881/aa9332}}.

\bibitem{Andersen2002}
I.~Andersen, A.~Szymkowiak, C.~Rasmussen, L.~Hanson, J.~Marstrand, H.~Larsson,
  L.~Hansen, Perfusion quantification using {Gaussian} process deconvolution,
  Magn. Reson. Med. 48~(2)  351--361.
\newblock \href {http://dx.doi.org/10.1002/mrm.10213}
  {\path{doi:10.1002/mrm.10213}}.

\bibitem{Christen2016}
J.~A. Christen, B.~Sans{\'{o}}, M.~Santana-Cibrian, J.~X.
  Velasco-Hern{\'{a}}ndez, Bayesian deconvolution of oil well test data using
  {Gaussian} processes, J. Appl. Stat. 43~(4) (2016) 721--737.
\newblock \href {http://dx.doi.org/10.1080/02664763.2015.1077374}
  {\path{doi:10.1080/02664763.2015.1077374}}.

\bibitem{Hunt1977}
B.~Hunt, Bayesian methods in nonlinear digital image restoration, IEEE Trans.
  Comput. C-26~(3) (1977) 219--229.
\newblock \href {http://dx.doi.org/10.1109/TC.1977.1674810}
  {\path{doi:10.1109/TC.1977.1674810}}.

\bibitem{CowanUnfolding}
G.~Cowan,
  \href{http://www.ippp.dur.ac.uk/old/Workshops/02/statistics/proceedings/cowan.pdf}{A
  survey of unfolding methods for particle physics}, in: Proc. Conference on
  Advanced Statistical Techniques in Particle Physics, Durham, England, 2002,
  pp. 248--257.
\newline\urlprefix\url{http://www.ippp.dur.ac.uk/old/Workshops/02/statistics/proceedings/cowan.pdf}

\bibitem{Blobel}
V.~Blobel, Unfolding methods in particle physics, in: Proc. PHYSTAT 2011
  Workshop on Statistical Issues Related to Discovery Claims in Search
  Experiments and Unfolding, CERN, Geneva, Switzerland, 2011, pp. 240--251.
\newblock \href {http://dx.doi.org/10.5170/CERN-2011-006.240}
  {\path{doi:10.5170/CERN-2011-006.240}}.

\bibitem{Spano:2013nca}
F.~Span\`o, {Unfolding in particle physics: a window on solving inverse
  problems}, EPJ Web Conf. 55 (2013) 03002.
\newblock \href {http://dx.doi.org/10.1051/epjconf/20135503002}
  {\path{doi:10.1051/epjconf/20135503002}}.

\bibitem{Rasmussen2006}
C.~E. Rasmussen, C.~K.~I. Williams, {{Gaussian} Processes for Machine
  Learning}, The MIT Press, 2006.

\bibitem{Cowan1998}
G.~Cowan, {Statistical Data Analysis}, Oxford University Press, 1998.

\bibitem{Tikhonov1952}
A.~Tikhonov, {On the solution of ill-posed problems and the method of
  regularization}, Mat. Sb. 151~(3) (1963) 501--504.

\bibitem{Phillips1962}
D.~L. Phillips, A technique for the numerical solution of certain integral
  equations of the first kind, J. ACM 9~(1) (1962) 84--97.
\newblock \href {http://dx.doi.org/10.1145/321105.321114}
  {\path{doi:10.1145/321105.321114}}.

\bibitem{Schmelling1994}
M.~Schmelling, The method of reduced cross-entropy: A general approach to
  unfold probability distributions, Nucl. Inst. Methods Phys. Res. A 340~(2)
  (1994) 400--412.
\newblock \href {http://dx.doi.org/10.1016/0168-9002(94)90119-8}
  {\path{doi:10.1016/0168-9002(94)90119-8}}.

\bibitem{Zech2013}
G.~Zech, Iterative unfolding with the {Richardson-Lucy} algorithm, Nucl. Inst.
  Methods Phys. Res. A 716 (2013) 1--9.
\newblock \href {http://arxiv.org/abs/1210.5177} {\path{arXiv:1210.5177}},
  \href {http://dx.doi.org/10.1016/j.nima.2013.03.026}
  {\path{doi:10.1016/j.nima.2013.03.026}}.

\bibitem{DAgostini1995}
G.~D'Agostini, {A multidimensional unfolding method based on Bayes' theorem},
  Nucl. Inst. Methods Phys. Res. A 362 (1995) 487--498.
\newblock \href {http://dx.doi.org/10.1016/0168-9002(95)00274-X}
  {\path{doi:10.1016/0168-9002(95)00274-X}}.

\bibitem{Choudalakis2012}
G.~Choudalakis, Fully {Bayesian} unfolding (2012).
\newblock \href {http://arxiv.org/abs/1201.4612} {\path{arXiv:1201.4612}}.

\bibitem{scikit-learn}
F.~Pedregosa, G.~Varoquaux, A.~Gramfort, V.~Michel, B.~Thirion, O.~Grisel,
  M.~Blondel, P.~Prettenhofer, R.~Weiss, V.~Dubourg, J.~Vanderplas, A.~Passos,
  D.~Cournapeau, M.~Brucher, M.~Perrot, E.~Duchesnay, Scikit-learn: Machine
  learning in {P}ython, Journal of Machine Learning Research 12 (2011)
  2825--2830.

\bibitem{Frate2017}
M.~Frate, K.~Cranmer, S.~Kalia, A.~Vandenberg-Rodes, D.~Whiteson, Modeling
  smooth backgrounds and generic localized signals with {Gaussian} processes
  (2017).
\newblock \href {http://arxiv.org/abs/1709.05681} {\path{arXiv:1709.05681}}.

\bibitem{Williams2007}
O.~Williams, A.~Fitzgibbon,
  \href{https://www.microsoft.com/en-us/research/publication/gaussian-process-implicit-surfaces-2/}{{Gaussian}
  process implicit surfaces}, in: Proc. Gaussian Processes in Practice, 2007,
  pp. 1--4.
\newline\urlprefix\url{https://www.microsoft.com/en-us/research/publication/gaussian-process-implicit-surfaces-2/}

\bibitem{Wahba1990}
G.~Wahba, Spline Models for Observational Data, Society for Industrial and
  Applied Mathematics, 1990.
\newblock \href {http://dx.doi.org/10.1137/1.9781611970128}
  {\path{doi:10.1137/1.9781611970128}}.

\bibitem{github}
\url{https://github.com/adambozson/gp-unfold}.

\bibitem{1975CoPhC..10..343J}
F.~James, M.~Roos, {Minuit -- a system for function minimization and analysis
  of the parameter errors and correlations}, Comput. Phys. Commun. 10 (1975)
  343--367.
\newblock \href {http://dx.doi.org/10.1016/0010-4655(75)90039-9}
  {\path{doi:10.1016/0010-4655(75)90039-9}}.

\bibitem{iminuit}
\href{https://github.com/iminuit/iminuit}{{iminuit -- A Python interface to
  Minuit}}.
\newline\urlprefix\url{https://github.com/iminuit/iminuit}

\bibitem{Gibbs1997}
M.~N. Gibbs, Bayesian {Gaussian} processes for regression and classification,
  Ph.D. thesis, Univ. Cambridge (1997).

\bibitem{Aad2016}
G.~{Aad et al. (ATLAS Collaboration)}, {Measurements of top-quark pair
  differential cross-sections in the lepton+jets channel in pp collisions at
  $\sqrt{s}=8 \text{ TeV}$ using the ATLAS detector}, Eur. Phys. J. C 76 (2016)
  538.
\newblock \href {http://dx.doi.org/10.1140/epjc/s10052-016-4366-4}
  {\path{doi:10.1140/epjc/s10052-016-4366-4}}.

\bibitem{Aad2016-2}
G.~{Aad et al. (ATLAS Collaboration)}, {Measurement of the differential
  cross-section of highly boosted top quarks as a function of their transverse
  momentum in $\sqrt{s} =8 \text{ TeV}$ proton-proton collisions using the
  ATLAS detector}, Phys. Rev. D 93 (2016) 1--34.
\newblock \href {http://dx.doi.org/10.1103/PhysRevD.93.032009}
  {\path{doi:10.1103/PhysRevD.93.032009}}.

\bibitem{shah2014student}
A.~Shah, A.~Wilson, Z.~Ghahramani, Student-t processes as alternatives to
  {G}aussian processes, in: Proc. Artificial Intelligence and Statistics, 2014,
  pp. 877--885.

\end{thebibliography}

\end{document}